\documentclass[
aip, jcp,
amsmath, amssymb,
reprint,
]{revtex4-1}

\usepackage{graphicx}

\usepackage{cases}
\usepackage{braket}
\usepackage{siunitx}

\usepackage{placeins}

\usepackage{hyperref}
\hypersetup{
  pdfpagemode=None, 
  pdfstartpage=1,
  pdfstartview=FitH,
  pdfmenubar=true,
  pdftoolbar=true,
  colorlinks = true,
  linkcolor=blue,
  citecolor=blue,
  bookmarksopen=false,
}

\newcommand{\unitsphere}[1]{\ensuremath{\mathrm{S}^{#1}}}
\newcommand{\SO}[1]{$\mathrm{SO}(#1)$}
\newcommand{\orientation}{\omega}

\renewcommand{\vec}{\boldsymbol}
\newcommand{\mat}{\mathbf}

\newcommand{\acc}{\mathcal{L}}
\newcommand{\eff}{\mathcal{E}}

\newcommand{\kB}{\ensuremath{k_{\mathrm{B}}}}

\newcommand{\mmax}{\mathrm{max}}

\newcommand{\mrot}{\mathrm{rot}}

\newcommand{\PAD}{\ensuremath{\mathrm{PAD}}}

\newcommand{\cofcl}{COFCl}

\begin{document}
\title{
Numerical evaluation of orientation averages and its application to molecular physics
}

\date{\today}

\author{Alexander Blech}
\author{Raoul M. M. Ebeling}
\author{Marec Heger}
\author{Christiane P. Koch}
\email{christiane.koch@fu-berlin.de}
\author{Daniel M. Reich}
\email{danreich@zedat.fu-berlin.de}
\affiliation{Dahlem Center for Complex Quantum Systems and Fachbereich Physik, Freie Universit\"at Berlin, Arnimallee 14, D-14195 Berlin, Germany}

\begin{abstract}
In molecular physics, it is often necessary to average over the orientation of molecules when calculating observables, in particular when modelling experiments in the liquid or gas phase. Evaluated in terms of Euler angles, this is closely related to integration over two- or three-dimensional unit spheres, a common problem discussed in numerical analysis. The computational cost of the integration depends significantly on the quadrature method, making the selection of an appropriate method crucial for the feasibility of simulations.
After reviewing several classes of spherical quadrature methods in terms of their efficiency and error distribution, we derive guidelines for choosing the best quadrature method for orientation averages and illustrate these with three examples from chiral molecule physics. 
While Gauss quadratures allow for achieving numerically exact integration for a wide range of applications, other methods offer advantages in specific circumstances.
Our guidelines can also be applied to higher-dimensional spherical domains and other geometries.
We also present a Python package providing a flexible interface to a variety of quadrature methods.
\end{abstract}

\maketitle

\section{Introduction}
The interaction of atoms and molecules with each other or with external fields critically depends on their relative orientation. Particularly in modelling liquid and gas phase experiments orientational averaging is required unless the orientation is precisely controlled. Since simulating molecular systems can be numerically expensive even for a single orientation, performing orientation averages, which involves numerous individual simulations, presents a significant computational challenge.
This problem is encountered across a broad range of molecular physics examples, including absorption spectroscopy \cite{Ma2006,Horsch2011},
photoelectron spectroscopy \cite{Wu2011,Lehmann2013,Beaulieu2016,Blanchet2021,Bloch2021,Artemyev2015,Tia2017,Demekhin2018,Nalin2021,Saribal2021,Tikhonov2022},
atom-molecule collisions \cite{Lombardi2018,Hutson2019},
molecules in external fields \cite{Tomza2013,Krems2018},
rotational dynamic simulations \cite{Leibscher2004,Leibscher2004a,Khodorkovsky2011,Gershnabel2018},
spin dynamics \cite{Kuprov2023},
and nonlinear optics \cite{Ayuso2019}.

Parameterized by the three Euler angles \cite{Goldstein1952,Edmonds1957,Zare1988}, orientation averaging is related to integrating over a unit sphere in four dimensions. System symmetries can reduce the dependence on one or two of these angles, simplifying the average to an integration over lower-dimensional spherical domains.
Quadrature methods, which approximate integrals through weighted sums over sampling points \cite{Dahlquist2008}, are well studied for spherical domains across a broad spectrum of scientific disciplines, including computer vision \cite{PerezSala2013,Marques2013}, geoscience \cite{Hesse2014,Gonzalez2010}, applied mathematics \cite{McLaren1963,Stroud1971,Atkinson1982,Cools1997,Lebedev1999,Graf2013,Beentjes2015}, radiation transport \cite{Manalo2015}, NMR spectroscopy \cite{Ebert1983,Alderman1985,Bak1997,Eden1998,Stoll2003,Eden2003,Afzali2021}, quantum chemistry \cite{Murray1993,Treutler1995,Daul1997}, and molecular physics \cite{Karney2007,Kuprov2023,Shimizu2023}.
While most studies focus on the mathematical properties of different quadrature schemes \cite{McLaren1963,Stroud1971,Atkinson1982,Cools1997,Daul1997,Lebedev1999,Karney2007,Gonzalez2010,Graf2013,Hesse2014,Marques2013,Beentjes2015,Kuprov2023}, others show benchmarks of selected methods in specific applications \cite{Ebert1983,Murray1993,Treutler1995,Bak1997,Eden1998,Eden2003,PerezSala2013,Beentjes2015,Saliba2019,Afzali2021,Shimizu2023}. Notably, comparative studies of three-angle quadratures are rare \cite{Eden1998,Eden2003,Karney2007,Shimizu2023}, with only Ref.~\onlinecite{Shimizu2023} providing a benchmark.

In molecular physics, numerical orientation averaging is not commonly addressed. Consequently, it is easy to overlook specialized and highly efficient averaging procedures that could substantially reduce computational effort compared to the often employed equidistant sampling. The absence of a comprehensive database for spherical quadratures further complicates method selection.

Here, we compare several methods for integrating over Euler angles, providing general recommendations based on the properties of the integrand and the desired precision to facilitate the choice of the best quadrature method. The difficulty of evaluating an orientation average largely depends on the number of Euler angles and the anisotropy of the integrand. We illustrate this through examples from state-of-the-art research in molecular physics. Specifically, we discuss how the symmetries of molecular systems and the properties of light-matter interactions manifest in the orientation dependence of angle-resolved photoelectron spectra and circular dichroism, both for randomly oriented and uniaxially oriented molecules. As an example for a three-angle orientation average we consider absorption circular dichroism with polarization-shaped laser pulses.

We restrict our discussion to Euler angles, the standard parameterization in physics, acknowledging potential drawbacks that can be addressed by alternative representations \cite{Mayerhofer2005}. One such alternative is the use of unit quaternions, commonly applied in computer graphics, aviation, and robotics \cite{Coutsias1999,Rapaport2004,LaValle2006,Karney2007,PerezSala2013}.

The remainder of this paper is organized as follows:
Section~\ref{averaging:euler_angle_integrals} formulates orientation averaging in terms of integrals over Euler angles, establishing the mathematical framework for the subsequent discussions.
In Section~\ref{averaging:measures}, we review and adapt selected measures to quantify the performance of numerical integration methods.
These performance measures are employed in Section~\ref{averaging:overview} to review various quadrature methods and provide general recommendations for method selection in Sec.~\ref{averaging:recommendations}.
Section~\ref{averaging:examples} applies these insights to three selected examples from the field of molecular physics, highlighting how the properties of the light-molecule interaction influence the choice of the quadrature method.
Finally, Sec.~\ref{averaging:python_package} introduces a software package offering a flexible, easy-to-use interface for a comprehensive collection of quadrature methods, facilitating their practical application.

\section{Orientation averages}
\label{averaging:euler_angle_integrals}
The orientation of an object is the rotation of its internal coordinate system with respect to the coordinate system of the observer, usually called the laboratory frame, assuming both frames are centered at a mutual origin. This rotation can be represented by a rotation matrix, which is an element of \SO{3}, the special orthogonal group. The average over all orientations in three-dimensional space is thus an average over \SO{3} \cite{Coutsias1999,Karney2007}.
Points in \SO{3} correspond to pairs of antipodal points on the unit sphere in four dimensions, \unitsphere{3}. Hence, it is possible to use \unitsphere{3} quadrature methods for orientation averaging.
Three real numbers are necessary to parameterize the rotation group. A common choice in physics are Euler angles $\orientation=(\alpha,\beta,\gamma)$ \cite{Goldstein1952,Edmonds1957,Zare1988}, which decompose an arbitrary rotation into three successive elemental rotations. Here, we employ the $zyz$-convention, such that $\alpha$ is the angle of rotation about the internal $z$-axis and $\gamma$ corresponds to a rotation about the laboratory-frame $z$-axis.

Using Euler angles, the orientation average of a function $f(\orientation)$ is given by the integral
\begin{align}
    I_3 = \int_0^{2\pi} \int_0^{\pi} \int_0^{2\pi} \sin(\beta) f(\alpha, \beta, \gamma) P(\alpha, \beta, \gamma)
    \,\mathrm{d}\alpha \mathrm{d}\beta \mathrm{d}\gamma
    \;.
    \label{eq:averaging:euler_angle_integral}
\end{align}
The probability density $P(\orientation)$ allows to individually weight different orientations. For example, in molecular physics $P(\orientation)$ can parameterize the orientational state of the molecule, as detailed in Appendix~\ref{averaging:appendix:euler_angle_distribution_and_rotational_states}.
In the following, we refer to $F(\orientation)\equiv f(\orientation)P(\orientation)$ as the integrand of Eq.~(\ref{eq:averaging:euler_angle_integral}).

If $F(\orientation)$ has azimuthal symmetries, it may not depend on the first or third Euler angle. In this case, the orientation average becomes the integral over a two-dimensional unit sphere \unitsphere{2}. An example is ionization of molecules with circularly polarized pulses, see Secs.~\ref{examples:rempi} and \ref{examples:cofcl}, and Appendix~\ref{averaging:appendix:simplified_averages}.

Equation~(\ref{eq:averaging:euler_angle_integral}) is composed of two different types of integrals: the integrals over the first and third Euler angle are integrals of the function $F(\phi)$ on a circle.
The integral over the second Euler angle is a polar integral, which can be expressed in two equivalent forms:
\begin{subequations}
\label{eq:averaging:polar_integral}
\begin{align}
I_\beta
&= \int_0^{\pi} \sin(\beta) F(\beta) 
\,\mathrm{d}\beta
\label{eq:averaging:polar_integral_theta}
\;, \\
&= \int_{-1}^{1} F(\arccos(z)) 
\,\mathrm{d}z
\;.
\label{eq:averaging:polar_integral_cos}
\end{align}
\end{subequations}
Although these two forms of the polar integral are equivalent, it makes a significant difference, whether a numerical integration method is applied to Eq.~(\ref{eq:averaging:polar_integral_theta}) or (\ref{eq:averaging:polar_integral_cos}).

\section{Performance measures for quadrature methods}
\label{averaging:measures}
A number of approaches exist in the literature for quantifying the performance of a quadrature method.
For instance, a priori estimates of the quadrature error, such as the Euler-MacLaurin formula \cite{Stoer2002}, provide the scaling of the error with the number of sampling points.
In one-dimensional quadratures, equally spaced sampling schemes are in most cases less efficient than Gauss quadratures with non-equidistant sampling points \cite{Davis1984,Dahlquist2008,Trefethen2013}. Conversely, uniform sampling is usually found to be superior on spherical domains \cite{Hesse2014}. 
Hence, spherical quadratures are commonly rated by their distribution of sampling points, quantified in terms of the covering radius \cite{Sloane2003,Karney2007,Hesse2014} and similar measures from the theory of covering codes (see Refs.~\onlinecite{Hifi2009,Conway2010} and references therein).
Another figure of merit is the number of polynomials that a quadrature method is able to integrate exactly. This is often associated with the "order" or "degree" of the quadrature method \cite{Lebedev1975,Cools1997,Hesse2014,Beentjes2015} and has been used by Ref.~\onlinecite{McLaren1963} to define a measure of efficiency.

In the following, we briefly review the concepts of polynomial exactness and efficiency and rigorously adapt them to the type of integrals relevant for orientation averaging. Although these concepts are important for understanding the asymptotic convergence properties of a quadrature method, they do not provide insights about the behaviour of the quadrature error in a specific application. To formulate recommendations for choosing the optimal method based on the properties of the integrand, we additionally adapt the idea of rank profiles from Ref.~\onlinecite{Eden1998}.

\subsection{Rank, exactness, and optimality}
Using the elements of Wigner D-matrices, $D^{l}_{mm'}(\orientation)$, as an orthogonal basis for orientation space \cite{Zare1988}, we represent the integrand $F(\orientation)$ in terms of a series expansion,
\begin{align}
F(\orientation) = \sum_l \sum_{m=-l}^l \sum_{m'=-l}^l
F_{lmm'} \, D^{l}_{mm'}(\orientation)
\;,
\label{eq:averaging:series_expansion}
\end{align}
where $F_{lmm'}$ are complex coefficients.
If the integrand does not depend on the first or the third Euler angle, then the Wigner D-matrix elements reduce to spherical harmonics, which provide an orthonormal basis \cite{Zare1988} on \unitsphere{2}.
We refer to $l$ as the \emph{rank} of the basis function. Accordingly, we define the \emph{maximum rank} of the integrand $F(\orientation)$ as the maximum value of $l$ beyond which the expansion coefficients $F_{lmm'}$ are negligibly small, e.g. smaller than the desired precision.
The rank of the basis functions encodes their anisotropy. Generally, an increasing number of sampling points are necessary to represent functions of increasing rank. Hence, numerical evaluation of the orientation average becomes particularly challenging if the integrand has high maximum rank.
This is the case for integrands with sharp features or discontinuities. In contrast, smooth integrands have smoothly and rapidly decaying expansion coefficients, making the integration less demanding.
As a result, the behavior of the expansion coefficients and the maximum rank of the integrand are good figures of merit for the expected numerical cost and the choice of the averaging method. We discuss this in more detail in the following sections.

Certain types of quadrature methods exactly integrate basis functions up to a given rank. This is referred to as the degree of precision \cite{Krylov1962,Ranga1994,Dryanov2006,Hesse2014}, degree of exactness \cite{Cools1997,Graf2013,Hesse2014,Gil2019} or algebraic order of accuracy \cite{Lebedev1995,Lebedev1999,Trefethen2008,Hesse2014}, sometimes simply termed order \cite{Lebedev1975,Lebedev1977,Eden1998,Hesse2014,Beentjes2015} or degree \cite{Cools1997,Hesse2014}.
However, the term "order" is not employed consistently across methods and fields, and is sometimes used interchangeably with the number of sampling points \cite{Beentjes2015}.

Here we adopt the term \emph{degree} as short-hand notation for the degree of exactness of a quadrature rule and denote it with $\acc$. A quadrature method has degree $\acc$ if it yields the exact value for the integral of $D^{l}_{mm'}(\orientation)$ for all $l\leq\acc$.
Quadrature rules which are inversion symmetric exactly integrate all Wigner D-matrix elements of odd rank to zero. In this case, the degree is the maximum even $l$ that fulfills the definition above.
A quadrature with degree equal to the maximum rank of the integrand or higher yields the numerically exact result of an orientation average, i.e., down to round-off errors due to machine precision.

A quadrature method is termed \emph{optimal} if it requires the minimum number of sampling points to achieve a given degree \cite{Hesse2014,Graf2013}.
Optimality is hard to prove and few examples exist, most notably Gauss-Legendre quadrature on the interval $[-1,1]$ \cite{Trefethen2013}. On \unitsphere{2}, evidence exists that Lebedev-Laikov quadrature is optimal \cite{Hesse2014}.
Nevertheless, many conjectured optimal quadrature schemes on \unitsphere{2} and \SO{3} are reported in the literature \cite{Sloane2003,Karney2007,Ahrens2009,Womersley2009,Mamone2010,Graf2011,Stepan2020,Popov2023}.

\subsection{McLaren efficiency}
A measure of a quadrature's efficiency is the ratio of its degree, $\acc$, over the number of sample point coordinates plus the number of weights, which constitute the quadrature rule's degrees of freedom \cite{McLaren1963}. Originally introduced by McLaren for quadratures on \unitsphere{2}, it can also be defined for one- and three-dimensional quadrature rules,
\begin{align}
\eff =
\begin{cases}
	\displaystyle
    \frac{\acc+1}{2n} \quad
    &\text{in one dimension,} \;,
    \\[1em]
    \displaystyle
    \frac{(\acc+1)^2}{3n} \quad
    &\text{on \unitsphere{2} (from Ref.~\onlinecite{McLaren1963})} \;,
    \\[1em]
    \displaystyle
    \frac{(1+\acc) (1+2\acc) (3+2\acc)}{12n} \quad
    &\text{on \SO{3}} \;,
\end{cases}
\label{eq:averaging:efficiency}
\end{align}
where $n$ is the number of sampling points.
The higher the efficiency, the fewer sampling points are necessary to reach a given degree of exactness. In that sense, the McLaren efficiency can be used to estimate the computational cost to perform an exact orientation average numerically.

Upper bounds for the efficiency exist for $\eff>1$ \cite{Delsarte1977,Steinerberger2021}. However, there exists evidence that optimality is equivalent to an asymptotic efficiency $\eff=1$, i.e. for $\acc\rightarrow\infty$. It is the case for the polar integral from Eq.~(\ref{eq:averaging:polar_integral_cos}), since Gauss-Legendre quadrature is optimal and has asymptotic efficiency $\eff=1$. The generalization of Gauss quadrature to spherical domains (cf. Sec.~\ref{averaging:overview:gauss}) asymptotically has $\eff=1$ and is believed to be optimal \cite{Hesse2014,Beentjes2015,Shimizu2023}. Thus, and due to a lack of quadratures with $\eff>1$ for $\acc\rightarrow\infty$, it is conjectured that optimal quadrature methods have asymptotic efficiency $\eff=1$ also in higher dimensions \cite{Hesse2014}.
Under this assumption, determining $n$ from Eq.~\eqref{eq:averaging:efficiency} with $\eff=1$ can serve as an estimate of the numerical cost to achieve a given $\acc$.

\subsection{Rank profiles}
Considering the orientation average of individual terms in the series expansion according to Eq.~(\ref{eq:averaging:series_expansion}) leads to the concept of \emph{rank profiles}, originally introduced by Ref.~\onlinecite{Eden1998}.
Due to the orthogonality of the Wigner D-matrix elements, only the $l=0$ term survives an exact orientation average, whereas terms with $l>0$ vanish.
In other words, an orientation average isolates the isotropic contribution of the integrand.
Accordingly, one can benchmark an averaging method by its effectiveness in removing anisotropic components of the integrand.

This can be quantified in terms of the sampling moments \cite{Eden1998},
\begin{align}
\sigma_{lmm'} = \frac{1}{8\pi^2} \int D^{l}_{mm'}(\orientation) \,\mathrm{d}\orientation
= \delta_{l0}
\;.
\label{eq:samplingmoments}
\end{align}
The quadrature error of a method is given by its approximate values of the sampling moments times the corresponding expansion coefficients of the integrand, $F_{lmm'}$.

To simplify the discussion, one can define the accumulated sampling moments,
\begin{align}
\Sigma_l =
\frac{1}{(2l+1)} \left(\sum_{m,m'} |\sigma_{lmm'}|^2\right)^{1/2}
\;.
\end{align}
For a given integration method, $\Sigma_l$ as a function of $l$ is called the \emph{rank profile of the method} \cite{Eden1998}.
It encodes the distribution of a method's quadrature error over the anisotropic basis functions on the integration domain.
In particular, methods with degree $\acc$ have zero sampling moments for $0<l\leq\acc$.

Selecting the optimal quadrature method for a specific application can be formulated as minimizing the overlap of the method's rank profile with the \emph{rank profile of integrand},
given by the accumulated expansion coefficients \cite{Eden1998},
\begin{align}
\Phi_l =
\frac{1}{(2l+1)} \left(\sum_{m,m'} |F_{lmm'}|^2\right)^{1/2}
\;.
\label{eq:averaging:rank_profile_integrand_three_angle}
\end{align}
Although calculating the rank profile of an integrand requires more numerical effort than the orientation average itself, in molecular physics, the close connection between the integrand's rank and angular momentum enables predictions about the rank profile by physical intuition or analytical estimates (see Appendix~\ref{averaging:appendix:euler_angle_distribution_and_rotational_states}).
In Sec.~\ref{averaging:examples}, we demonstrate this with photoelectron angular distributions, predicting their rank profiles based on the light-molecule interaction and relevant molecular states.
Rank profiles can be defined for arbitrary domains. We provide a general definition and additional examples in Appendix~\ref{averaging:appendix:rank_profiles}.

Analyzing the sampling moments $\sigma_{lmm'}$ and expansion coefficients $F_{lmm'}$ with respect to the azimuthal indices $m$ and $m'$ provides an even more detailed view onto the symmetry properties of the quadrature error and the integrand (see Appendix~\ref{averaging:appendix:method_rank_profiles}).

\section{Comparison of quadrature methods}
\label{averaging:overview}
Spherical quadratures may be broadly categorized into five main groups:
spherical Gauss quadratures, spherical Chebyshev quadratures, near-uniform spherical coverings, product quadratures and Monte-Carlo methods.
In the following, we review and compare these approaches for two- and three-angle quadratures. To formulate general recommendations for the selection of the best quadrature method (summarized in Sec.~\ref{averaging:recommendations}), we calculated their McLaren efficiency, and rank profiles.
A chart of the efficiency and exemplary rank profiles of representative quadrature methods from each category are provided in Appendices~\ref{averaging:appendix:efficiency_chart} and \ref{averaging:appendix:method_rank_profiles}, respectively.

\subsection{Spherical Gauss quadratures}
\label{averaging:overview:gauss}
Spherical Gauss quadratures (as coined by Ref.~\onlinecite{Eden1998}) generalize one-dimensional Gauss quadratures to \unitsphere{2} or \SO{3} by approximating the integrand by Wigner D-matrix elements or spherical polynomials. Sampling points and weights are obtained from the condition to achieve a given degree $\acc$ \cite{Press1986,Eden1998,Hesse2014,Beentjes2015}.
Solving the resulting large systems of equations is challenging.
Sobolev reduced this complexity on \unitsphere{2} by looking for quadrature rules that are invariant with respect to finite subgroups of \SO{3} \cite{Sobolev1962}. This approach has been successfully applied to construct spherical Gauss quadratures with sampling points distributed according to different point groups \cite{Lebedev1975,Popov1995,Daul1997,Stoyanova1997,Ahrens2009}.
On \unitsphere{2}, a notable example is Lebedev-Laikov quadrature \cite{Lebedev1975,Lebedev1976,Lebedev1977,Lebedev1992,Lebedev1995,Lebedev1999}, which is invariant with respect to the octahedral point group and is available up to $\acc=131$.
It is widely used in computational chemistry\cite{Murray1993,Treutler1995,Daul1997,Koch2001,Wang2003,Furuhama2004} and often regarded as the best-performing method on \unitsphere{2} for smooth integrands \cite{Eden1998,Eden2003,Beentjes2015,Afzali2021,Shimizu2023}.
An improved version has been proposed by Ref.~\onlinecite{Stevensson2006}. \unitsphere{2} Gauss quadratures with icosahedral symmetry  have been reported \cite{Ahrens2009,Ahrens2012} with $\acc\leq210$ and Ref.~\onlinecite{Daul1997} outlines a procedure to construct spherical Gauss methods invariant with respect to arbitrary point groups.

Alternatively, spherical Gauss quadratures can be obtained by numerical optimization of the sampling points and weights \cite{Ahrens2009,Sloan2009,Ahrens2012,Graf2013,Dai2020,Voloshchenko2020,Dai2022,Popov2023}. This strategy is particularly useful for constructing quadratures of high degree where analytical derivations become increasingly difficult. These numerically optimized methods retain the properties of Gauss quadratures, such as degree and symmetries, within finite numerical precision. They are expected to offer better performance and potentially achieve slightly higher efficiency compared to their exact counterparts \cite{Stevensson2006,Dai2022,Shimizu2023}.
This is confirmed by Fig.~\ref{averaging:appendix:efficiency_chart} for the \unitsphere{2} and \SO{3} methods from Ref.~\onlinecite{Graf2013}, showing efficiency $\eff>1$ for some values of $\acc\gtrsim10$.
Notably, the quadratures from Ref.~\onlinecite{Graf2013} are the only known \SO{3} Gauss quadratures. They are available for $\acc\leq14$.

All known spherical Gauss quadratures are conjectured to be optimal and have asymptotic efficiency $\eff=1$ (for $\acc\rightarrow\infty$) \cite{Hesse2014,Beentjes2015}.
Rank profiles of Gauss quadratures have the typical characteristics of a method with high degree: sampling moments peak sharply at $l=\acc+1$ and stay comparatively large for higher $l$. Consequently, these methods acquire a large quadrature error, as soon as the maximum rank of the integrand exceeds the method's degree.

\subsection{Spherical Chebyshev quadratures}
\label{averaging:overview:chebyshev}
Chebyshev quadratures are constructed to achieve a specified degree $\acc$ with uniform quadrature weights, resulting in a near-uniform distribution of points.
On $d$-dimensional unit spheres, \unitsphere{d}, they are called spherical designs \cite{Delsarte1977,Sloane2003} (cf. Ref.~\onlinecite{Bannai2009} for a review).
Spherical Chebyshev quadratures for three-angle averages can be constructed from inversion-symmetric spherical designs on \unitsphere{3}, by using pairs of antipodal points on \unitsphere{3} as sampling points in \SO{3} \cite{Graf2013}.

Spherical Chebyshev quadratures exist for any degree $\acc$ \cite{Seymour1984}.
Their efficiency has an upper bound of $\eff<5/4$ for $\acc<29$, approaching $\eff<4/3$ for $\acc\rightarrow\infty$ \cite{Delsarte1977}. However, this bound is never reached on \unitsphere{2} except for $\acc=1,2,3$ and $5$, for which $\eff\leq1$ \cite{Bannai1979,Reimer1992}.
Reference~\onlinecite{Chen2011} gives an lower bound on efficiency, $\eff=1/3$, for \unitsphere{2} spherical designs with $L\leq100$. Although no proof exists that spherical Chebyshev quadratures with $\eff\approx1$ do not exist beyond $\acc=5$, the known spherical designs with $\acc>5$ rapidly converge to $\eff\lesssim2/3$ for $\acc\rightarrow\infty$.
This asymptotic efficiency has not yet been overcome, either by analytically derived methods nor by numerical optimization of the quadrature points \cite{Sloan2009,Graf2011}.
Thus, it is conjectured that $\eff=2/3$ is the maximum asymptotic efficiency of spherical Chebyshev quadratures \cite{Karney2007,Womersley2009,Graf2011}, meaning they require about \SI{50}{\%} more sampling points than spherical Gauss quadratures of the same degree.
Spherical Chebyshev quadratures with the conjectured minimal number of points currently exist up to $\acc\leq181$ on \unitsphere{2} \cite{Womersley2009,Graf2011} and $\acc\leq23$ on \SO{3} \cite{Karney2007,Womersley2009,Graf2013}.

The rank profiles of spherical Chebyshev quadratures resemble those of spherical Gauss quadratures, yet exhibit broader initial peak and smaller sampling moments for high $l$. This suggests that Chebyshev quadratures may yield smaller errors than Gauss methods if the maximum rank of the integrand exceeds $\acc$.

\subsection{Uniform spherical coverings}
\label{averaging:overview:coverings}
Quadrature methods based on uniform sampling of the unit sphere or orientation space are known as spherical coverings \cite{Karney2007,Womersley2009} or spherical codes \cite{Sloane2003,Hifi2009,Conway2010}.
Like Chebyshev methods, spherical coverings have uniform weights. Although individual spherical coverings can be classified as spherical designs, they do not have $\acc>0$ by construction.
Nevertheless, their properties closely resemble those of spherical Chebyshev quadratures.

Because perfectly uniform distribution of more than 20 points on \unitsphere{2} is impossible \cite{Bannai1979,Ahrens2012,Mamone2010}, a variety of approaches exist to find near-uniform distributions.
The REPULSION method from Ref.~\onlinecite{Bak1997} distributes points as electrostatically charged particles.
Mapping Fibonacci lattices onto the surface of the unit sphere gives rise to Fibonacci spheres \cite{Gonzalez2010,Marques2013}.
A higher-dimensional variant of Fibonacci spheres is the ZCW method, named after Zaremba, Conroy, and Wolfsberg \cite{Zaremba1966,Conroy1967,Cheng1973}, which uses Fibonacci lattices to construct quadrature methods for integrals in arbitrary dimensions.
Other approaches include equal area or equal volume partitions of the integration domain \cite{Sloane2003,Womersley2009,Mamone2010} or, generally speaking, using Voronoi tessellation or Delaunay triangulation \cite{Ebert1983,Stoll2003,Aurenhammer2013,Karney2007}.
These can also be constructed with certain symmetry (e.g. the octahedral EasySpin grid from Ref.~\onlinecite{Stoll2003}) or applied adaptively \cite{Ebert1983,Stoll2003,Kuprov2023}.

Since all spherical coverings share the same underlying principle, their quadrature errors behave very similarly. Their rank profiles closely resemble those of spherical Chebyshev methods with a comparable number of sampling points. However, because spherical coverings have $\acc=0$, all sampling moments are non-zero, leading to slightly larger quadrature errors than Chebyshev methods and a different error-scaling with the sampling density.

\subsection{Tensor product quadratures}
\label{averaging:overview:products}
Using Fubini's theorem \cite{Fubini1907}, two- and three-angle quadrature schemes can be constructed by combining multiple lower-dimensional quadrature schemes.
The resulting product quadrature inherits properties from the methods it is composed of.
A common issue with product rules is the clustering of sampling points with small weights at the poles, leading to inefficient coverage of the integration domain. While this could be beneficial if the integrand has sharp features near the poles, such bias is usually unnecessary since the choice of poles in orientation averaging is arbitrary. Moreover, non-uniform sampling induces sensitivity to arbitrary rotations of the integrand, reducing robustness compared to more uniform sampling.
However, the individual quadratures of a product method can be tuned for the different Euler angles individually to best exploit the integrand's properties or potential symmetries.

Popular examples of product methods are combinations of several low-degree composite equidistant quadratures, known as step methods \cite{Eden1998}. These have recently been used to calculate orientation-averaged photoelectron spectra\cite{Artemyev2015,Tia2017,Muller2018,Demekhin2018,Demekhin2019,Muller2020,Beaulieu2016,Blanchet2021,Bloch2021,Tikhonov2022}.
Step methods yield smaller sampling moments and quadrature errors with equidistant sampling in $\beta$, Eq.~(\ref{eq:averaging:polar_integral_theta}), instead of equidistant sampling in $\cos\beta$, Eq.~(\ref{eq:averaging:polar_integral_cos}).
Despite having $\acc=0$, their small sampling moments allow step methods to compete with spherical Gauss or Chebyshev quadratures if the rank of the integrand is much larger than their degree.

Product rules with arbitrary $\acc$ can, e.g., be constructed by combining the composite trapezoidal method for $\alpha$ and $\gamma$ with Gauss-Legendre quadrature for $\beta$.
Note that for Gauss product quadratures to have non-zero degree with respect to integrating spherical harmonics or Wigner-D matrix elements it is critical to evaluate the $\beta$-integral in the form of Eq.~(\ref{eq:averaging:polar_integral_cos}).
They have efficiency $\eff\lesssim2/3$ for all $\acc$, with the same asymptotic efficiency as spherical Chebyshev quadratures,  \cite{Hesse2014,Ahrens2009,Beentjes2015} but lower efficiency for low degrees (see Fig.~\ref{averaging:appendix:efficiency_chart} in the appendix).
The rank profiles of Gauss product methods show an isolated peak at $l=\acc+1$ with small sampling moments for several $l>\acc+1$. This indicates that their quadrature error may be smaller than the error of spherical Gauss methods if the maximum rank of the integrand is larger than $\acc$.

For three-angle averages, product rules can also combine spherical quadrature methods with one-dimensional methods.
Examples include the combination of the composite trapezoidal method with spherical designs (denoted T$\times$D) or with spherical Gauss methods (denoted T$\times$G) \cite{Shimizu2023}.
These hybrid product methods have the same efficiency as their underlying \unitsphere{2} methods.
In particular, the T$\times$G method has efficiency $\eff=1$ for $\acc\rightarrow\infty$, making it a promising extension of the spherical Gauss methods to the three-angle case.
The rank profiles of the T$\times$D and T$\times$L methods behave in the same way as the rank profiles of \SO{3} Chebyshev and Gauss quadratures, respectively.

\subsection{Monte-Carlo methods}
\label{averaging:overview:mc}
Monte-Carlo integration is based on random sampling of the integration domain. Although this becomes increasingly efficient for higher-dimensional integrals, it is outperformed on \unitsphere{2} and \SO{3} by the deterministic methods discussed earlier. Monte-Carlo methods generally have larger sampling moments, slower convergence by several orders of magnitude and less robustness than other spherical quadratures, in line with observations reported in the literature \cite{Karney2007,Beentjes2015}.
Notably, quaternion representation is superior than Euler angles for random sampling of orientation space, e.g. by Shoemake's method \cite{Shoemake1992}. Hence, quaternions find widespread use in Monte-Carlo molecular dynamic simulations \cite{LaValle2006,Karney2007,Tutunnikov2018}.

\section{Recommendations and guidelines}
\label{averaging:recommendations}
If the integrand's maximum rank is low enough to allow for exact orientation averaging with a quadrature of sufficiently high degree, the best choice is the method with the highest efficiency, as determined from the efficiency chart in Appendix~\ref{averaging:appendix:efficiency_chart}.
We exemplify this for photoelectron circular dichroism after multi-photon ionization of randomly oriented molecules in Sec.~\ref{examples:rempi}.

If the required degree exceeds the maximum available for Gauss methods, spherical Chebyshev quadratures offer an alternative, at the expense of lower efficiency. If the available maximum degree of spherical Chebyshev methods is still insufficient, Gauss product methods provide a way to construct quadrature schemes of arbitrary degree with similar efficiency but less robustness with respect to random rotations of the integrand.

For high-rank integrands where exact quadrature is too costly, or limited precision is sufficient, step methods and spherical coverings can yield smaller quadrature errors than high-degree methods with less sampling points.
Near-uniform spherical coverings are particularly promising due to their small sampling moments, as demonstrated in terms of photoelectron spectra from oriented molecules and absorption circular dichroism with elliptically polarized laser pulses in Secs.~\ref{examples:cofcl} and \ref{examples:cd}, respectively.

If the integrand exhibits symmetries, selecting a quadrature method with matching symmetry reduces the number of sampling points. Suitable methods can often be found among spherical Gauss quadratures or product methods.
Product methods can also be tailored to the integrand's properties to minimize the quadrature error. We provide evidence for the example of photoelectron spectra from oriented molecules in Sec.~\ref{examples:cofcl}. This flexibility is particularly useful if the integrand only depends weakly on individual Euler angles, e.g. if it exhibits approximate symmetries.

\section{Case studies}
\label{averaging:examples}
As discussed in Sec.~\ref{averaging:measures}, the difficulty of an orientation average depends on the number of relevant Euler angles and the integrand's rank profile.
In molecular physics these parameters are determined by the symmetry of the light-molecule interaction, the number of photons interacting with the molecule, and the initial orientational distribution of the molecular ensemble.
We exemplify this with the following three case studies.

For all examples, quadrature methods are benchmarked by the relative quadrature error,
\begin{align}
\epsilon(n) = \frac{| I(n) - I(n_{\mmax}) |}{I(n_{\mmax})}
\;,\label{eq:relative_quadrature_error}
\end{align}
where $I(n)$ is the numerical value of the orientation average obtained with $n$ sampling points and $n_{\mmax}$ denotes the maximum available number of sampling points for the corresponding method.

\subsection{Low rank, two angles: multi-photon ionization of randomly oriented molecules}
\label{examples:rempi}
\begin{figure*}
\includegraphics[width=\linewidth]{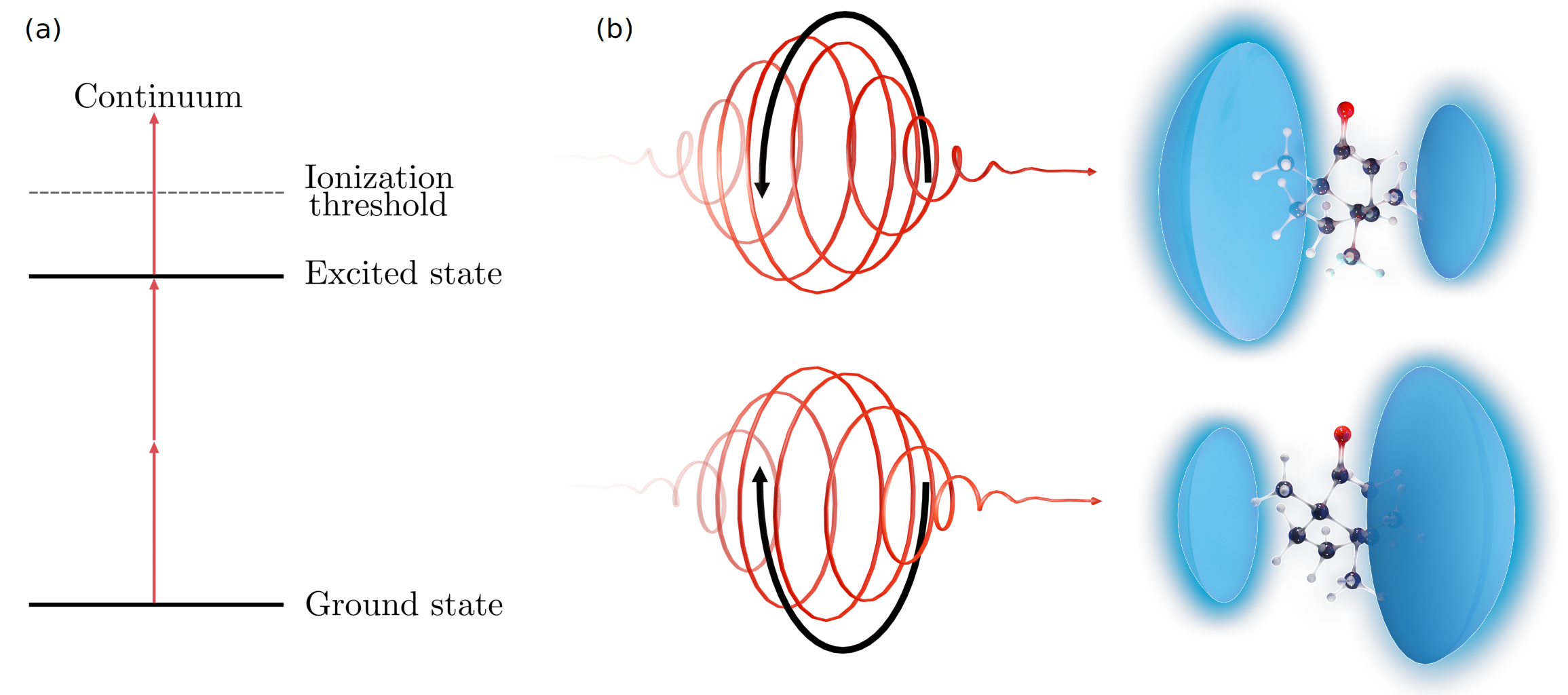}
\caption{
\textbf{(a)}
Excitation diagram for 2+1 REMPI featuring a two-photon ionization (photon energy $0.58$\,eV) from the molecular ground state of camphor to an intermediate state. Population in the intermediate state is ionized by a third photon into the continuum.
\textbf{(b)}
Illustration of photoelectron circular dichroism for camphor. The interaction of chiral molecules with circularly polarized light results in a forward-backwards asymmetry in the photoelectron angular distribution indicated by the blue shaded area. 
}
\label{fig:rempi}
\end{figure*}
\begin{figure}
\includegraphics[width=\linewidth]{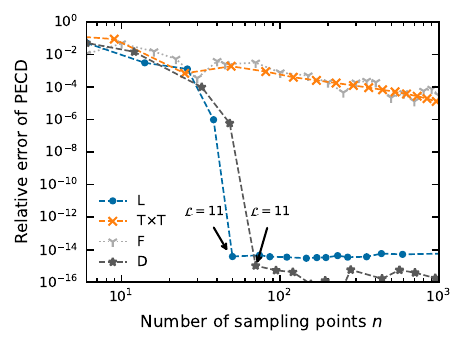}
\caption{
Relative quadrature error of PECD calculated with Eq.~(\ref{eq:relative_quadrature_error}) as resulting from the 2+1 REMPI process shown in Fig.~\ref{fig:rempi}(a) applied to randomly oriented molecules. We calculate PADs at \SI{0.58}{eV} photoelectron energy, averaged over the Euler angles $\beta$ and $\gamma$ for $\alpha=0$. The error is shown for the maximum PECD signal, corresponding to the forward direction ($\theta_k=0$).
Methods are Lebedev-Laikov quadrature (L), spherical designs from Ref.~\onlinecite{Womersley2009} (D), equidistant step methods with the same sampling density for $\beta$ and $\gamma$ (T$\times$T) and near-uniform spherical coverings by Fibonacci spheres (F).
The arrows indicate the degree $\acc$ for which the error of the Lebedev-Laikov method and the spherical designs reaches machine precision.
}
\label{fig:method_comparison}
\end{figure}
Our first example features cylindrical symmetry and an integrand of low rank. The cylindrical symmetry reduces the amount of relevant Euler angles to two, whereas the effect of low rank will be discussed below. We consider 2+1 resonance-enhanced multiphoton ionization (REMPI) of randomly oriented camphor as described by Ref.~\onlinecite{Goetz2017}. Briefly, simultaneous absorption of two photons excites an electron to an intermediate bound molecular state. From there, another photon is absorbed leading to ionization. This ionization scheme is shown in Fig.~\ref{fig:rempi}(a). Since camphor is chiral, the difference of the PADs obtained from left/right-circular polarization shows a forward-backward asymmetry with respect to the light propagation direction, known as photoelectron circular dichroism (PECD) \cite{Ritchie1976,Bowering2001,Lux2012} and illustrated in Fig.~\ref{fig:rempi}(b).

A PAD represents a three-dimensional momentum distribution of the emitted photoelectron that typically is expanded in terms of spherical harmonics for the angular part of the photoelectron momentum $\vec k=(k, \theta_k, \phi_k)$, here defined in the laboratory frame,
\begin{align}
\PAD(\vec k, \orientation) = \sum_{L M} b_{LM}(k, \orientation) \, Y_{LM}(\theta_k, \phi_k)
\;.
\label{eq:averaging:pad_expansion}
\end{align}
Linearly polarized pulses and circularly polarized pulses containing a sufficient number of cycles can be considered cylindrically symmetric around their polarization direction or the pulse propagation direction, respectively. By aligning one Euler rotation axis with this symmetry axis, the corresponding Euler angle simply rotates the PAD around this symmetry axis. Consequently, the orientation average can be reduced to an integral over the remaining two Euler angles. See Appendix~\ref{averaging:appendix:averaging_cylindrical_pads} for details.
The PECD signal is then obtained as
\begin{equation}
\text{PECD}(k,\theta_k)=\frac{\text{PAD}^-(k,\theta_k)-\text{PAD}^+(k,\theta_k)}{\text{PAD}^-(k,\theta_k)+\text{PAD}^+(k,\theta_k)},
\end{equation}
where the superscript denotes ionization with left- ($-$) and right- ($+$) circular polarization.

The relative error of the maximum PECD is shown in Fig.~\ref{fig:method_comparison} for various quadrature methods.
We observe that the equidistant product method as well as the uniform spherical coverings converge slowly compared to spherical Gauss and Chebyshev quadratures.
In particular, the error reaches machine precision for methods with $\acc\geq11$, corresponding to 50 sampling points for Lebedev-Laikov quadrature and 70 points for the spherical design from Ref.~\onlinecite{Womersley2009} due to their lower efficiency.

This behaviour can be directly related to the low rank this example features. In Appendix~\ref{averaging:appendix:rank_profile_of_pad}, we analytically derive the rank profile of laboratory-frame multi-photon PADs based on perturbation theory.
For an isotropic orientation average, the upper bound for the maximum rank of a PAD resulting from an interaction with $n_{\mathrm{ph}}$ photons is
\begin{align}
l_{\mmax}
\leq L_{\mmax} + 2 n_{\mathrm{ph}}
\leq 2 l_{k, \mmax} + 2 n_{\mathrm{ph}}
\;,
\label{eq:averaging:max_rank_of_pad}
\end{align}
where $L_{\mmax}$ is the maximum $L$ in the spherical harmonic expansion of the PAD, cf.~Eq.~(\ref{eq:averaging:pad_expansion}), and $l_{k, \mmax}$ is the highest partial wave quantum number of the photoelectron's final state.
The latter is determined by the angular momentum of the molecule's initial state and the selection rules. For a molecule without symmetry ($\mathrm{C}_{1}$) and for electric dipole transitions with circularly polarized light we have $l_{k, \mmax} = l_{0, \mmax} + n_{\mathrm{ph}}$, where $l_{0, \mmax}$ is the largest angular momentum quantum number of the initial state.
Due to three photons contributing to the overall ionization process, we obtain $L_{\mmax}=6$, restricting the maximum possible rank of the PAD to 12.
However, the calculations from Ref.~\onlinecite{Goetz2017} show that the intermediate state of the 2+1 REMPI has a strong p-character. Taking into account the additional photon that ionizes the electron from the intermediate state yields $l_{k, \mmax}\approx2$. With $n_{\mathrm{ph}}=3$, we thus obtain $l_{\mmax}=10$ for the maximum rank of resulting PADs. This predicts an error drop-off for quadratures with $\acc\geq10$ for a single PAD, which we confirmed numerically. Since PECD is a normalized difference of PADs, convergence for $\acc\geq10$ also holds. The aforementioned considerations explain the steep drop-off of the relative errors for spherical Gauss and Chebyshev methods in Fig~\ref{fig:method_comparison}. In contrast, spherical coverings and equidistant step methods lack such a feature.
Overall this example demonstrates the superior performance of methods with non-zero degree if the rank of the integrand is small. In such cases we advise to use a spherical Gauss quadrature due to their high efficiency. Even if the rank of the integrand is slightly misjudged these methods can still be expected to yield the lowest quadrature errors.

\FloatBarrier

\subsection{High rank, two angles: PECD of anisotropic molecular ensembles}
\label{examples:cofcl}
\begin{figure*}
\includegraphics{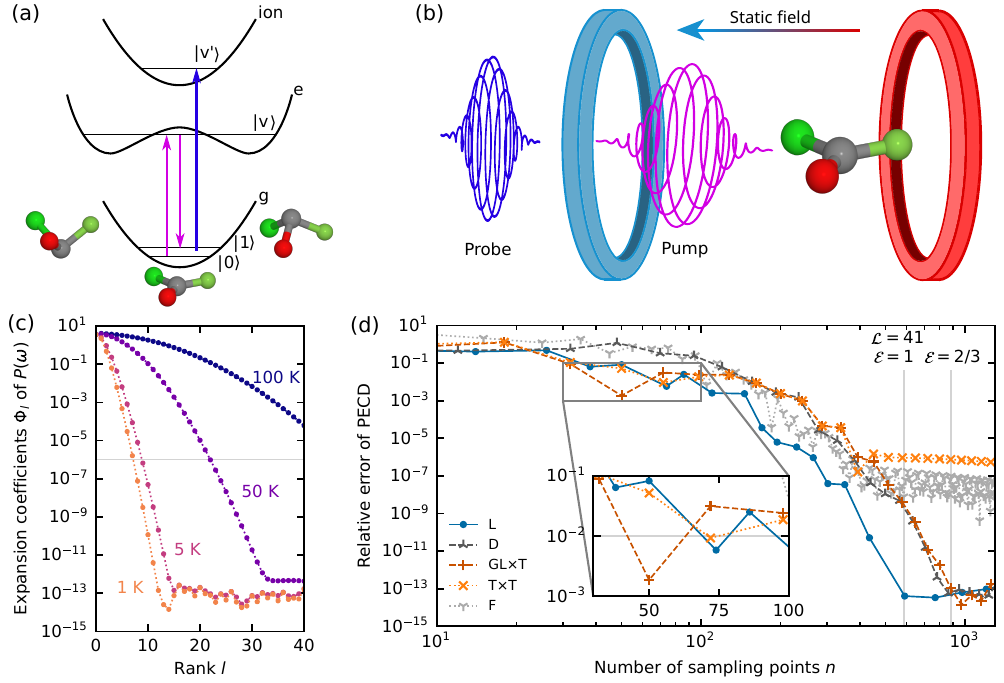}
\caption{
\textbf{(a)}
Exciting and probing a chiral vibrational wavepacket in planar COFCl. A Raman excitation (magenta arrows) creates a superposition of the two lowest out-of-plane vibrational states of the central C-atom via an electronically excited state with vibrational level $\ket{v}$, followed by one-photon ionization (blue) populating a vibrational level $\ket{v'}$ in the parent ion's ground state.
\textbf{(b)}
Visualization of the field configuration. Electric field coils generate the static field orienting the molecules. Pump (magenta) and probe (blue) pulses are circularly polarized in a plane perpendicular to this field.
\textbf{(c)}
Rank profile of the Euler angle distribution $P(\orientation)$ from Eq.~(\ref{eq:averaging:P_boltzmann}) for different temperatures obtained with the Lebedev-Laikov method of degree $\acc=131$.
The horizontal gray line indicates the value $10^{-6}$ used to determine the maximum rank of $P(\orientation)$.
\textbf{(d)}
Relative quadrature error of PECD calculated with Eq.~(\ref{eq:relative_quadrature_error}) for the process shown in (a) and (b) at \SI{6}{eV} photoelectron energy. The orientation average is weighted with the Euler angle distribution from Eq.~(\ref{eq:averaging:P_boltzmann}) at rotational temperature \SI{5}{K}.
The vertical gray lines indicate the number of sampling points needed to achieve degree $\acc=41$ for a method with efficiency $\eff=1$ and $\eff=2/3$.
The inset displays a zoom of the same data, with the horizontal gray line highlighting a relative error of \SI{1}{\percent}.
Methods are Lebedev-Laikov quadrature (L), the spherical designs from Ref.~\onlinecite{Womersley2009} (D), Gauss-Legendre product grids (GL$\times$T), equidistant step methods with the same sampling density for $\beta$ and $\gamma$ (T$\times$T) and near-uniform spherical coverings by Fibonacci spheres (F).
}
\label{fig:cofcl}
\end{figure*}
An anisotropic orientational distribution of molecules, can be modelled by weighting the orientation average in Eq.~(\ref{eq:averaging:euler_angle_integral}) with an anisotropic Euler angle distribution $P(\orientation)$. This increases the rank of the integrand by the maximum rank of $P(\orientation)$ --- and thus increases the numerical cost.
For molecular systems, the rank profile of the Euler angle distribution is linked to the orbital angular momentum of the molecule, as $P(\orientation)$ can be taken to be the absolute value of the rotational wavefunction.
Appendix~\ref{averaging:appendix:euler_angle_distribution_and_rotational_states} shows,
that the maximum rank of $P(\orientation)$ is twice the maximum rotational quantum number.

As an example, we consider the pump-probe experiment proposed in Ref.~\onlinecite{Tikhonov2022}. The process, illustrated in Fig.~\ref{fig:cofcl}(a), comprises three-photons:
a circularly polarized Raman pulse excites a chiral vibrational wavepacket in planar \cofcl{} molecules, which is then probed by one-photon ionization with a circularly polarized femtosecond pulse.
In contrast to the preceding example, the initial orientational distribution of the molecules is not uniform. A static electric field orients the molecules uniaxially along the pulse propagation direction, as displayed in Fig.~\ref{fig:cofcl}(b), preserving cylindrical symmetry of the light-molecule interaction and allowing for a two-angle orientation average.
Although \cofcl{} is a member of the $\mathrm{C}_{\mathrm{s}}$ symmetry group, the induced vibrational dynamics break the planar symmetry. Hence, no symmetries remain that would reduce the maximum rank of the PAD or simplify its dependence on the third Euler angle $\gamma$.
The chirality of the vibrational dynamics manifests itself in the PECD.

We estimate the rank of the Euler angle distribution by assuming thermal equilibrium \cite{Tikhonov2022},
\begin{align}
P(\orientation) = N(T) \, e^{- \frac{\vec E \cdot \mat R(\orientation) \cdot \vec d}{\kB T}}
\;,
\label{eq:averaging:P_boltzmann}
\end{align}
where $T$ is the rotational temperature of the ensemble, $N(T)$ is a normalization factor, $\vec E$ is the static electric field in the laboratory frame, $\vec d$ is the permanent dipole moment of the molecule in the molecular frame and $\mat R(\orientation)$ is the rotation matrix for the transformation between the two frames of reference.
Figure~\ref{fig:cofcl}(c) displays the rank profile of $P(\orientation)$ for different values of $T$.
Since $P(\orientation)$ essentially describes the distribution of the rotational states of the molecule, the rank profile of $P(\orientation)$ also follows a Boltzmann distribution, down to the precision limit resulting from round-off errors.
As $T$ approaches zero, the molecules become perfectly oriented and $P(\orientation)$ takes the form of a delta distribution, corresponding to a flat rank profile. For high temperatures, the orientation by the static field becomes less effective and the resulting rotational wavepacket remains more and more isotropic, decreasing the maximum rank of $P(\orientation)$. In the limit $T\rightarrow\infty$, we recover the case of uniformly oriented molecules, corresponding to $P(\orientation)\rightarrow\text{const.}$ which has maximum rank zero.

The maximum rank of the weighted integrand, $P(\orientation) \mathrm{PAD}(\vec k, \orientation)$, results from a product of two small numbers. The relevant threshold is the numerical precision, which is of the order of $10^{-12}$ in our calculations. Hence, we take the maximum rank of $P(\orientation)$ to be the maximum $l$ for which the sampling moments are above half of the precision limit, i.e., at $10^{-6}$.
For $T=\SI{5}{K}$, this yields $22$.
Compared to the example from Sec.~\ref{examples:rempi}, the rank of the orientation average is additionally increased by the fact that the ionization from the highest occupied molecular orbital of \cofcl{} populates partial waves up to $l_{k, \mmax}\approx 10$.
Thus, the PAD can in principle contain contributions from anisotropy parameters up to $L_{\mmax}\approx20$.
According to Eq.~(\ref{eq:averaging:max_rank_of_pad}), with three photons contributing to the ionization process we obtain $l_{\mmax} \approx 26$, yielding a maximum possible rank for the weighted orientation average of about $48$.

As a result, significantly more quadrature points are necessary to reach the same level of precision as in Sec.~\ref{examples:rempi}.
This is illustrated in Fig.~\ref{fig:cofcl}(d) showing the quadrature error of different methods as a function of the number of sampling points, $n$.
The error reaches machine precision for methods with degree $\acc\geq41$, which is only slightly less than the estimated maximum rank.
For spherical Gauss quadratures, which have efficiency $\eff\approx 1$, this corresponds to about $600$ sampling points, whereas for spherical designs and other methods with efficiency $\eff\approx2/3$, e.g. the Gauss product method, this amounts to more than $900$ quadrature points.
Furthermore, spherical Gauss methods outperform all other methods for quadrature errors of \SI{0.1}{\%} and below. In particular, the Lebedev-Laikov method exhibits the smallest error for $n\gtrsim100$, demonstrating a slight asymptotic advantage over the \unitsphere{2} Gauss quadrature from Ref.~\onlinecite{Graf2013} (not shown in the figure).

Methods with degree $\acc=0$ can achieve errors of the order of $10^{-6}$ with less than 1000 sampling points, too. Among these, the Fibonacci spheres provide the smallest error for $500\gtrsim n \gtrsim100$, even outperforming spherical designs and Gauss product methods.
In contrast to our other examples, the quadrature error of methods with $\acc=0$ shows a similar super-exponential scaling as for methods with $\acc>0$, up to $n\approx400$, before asymptotically showing a slow exponential decay (cf. Figs.~\ref{fig:method_comparison} and \ref{fig:CD_fenchone}(c)). Comparison with the rank profile of $P(\orientation)$ indicates that for a small number of sampling points the quadrature error is dominated by the rotational anisotropy.

Other error sources, such as the basis set error in the description of the molecule's electronic states, may not necessitate the evaluation of the orientation average with ultrahigh precision.
Also, the computational cost may render sufficiently large $n$ to exploit the advantage of spherical Gauss methods unfeasible.
As indicated in Sec.~\ref{averaging:overview}, spherical Gauss methods are not necessarily the optimal choice in this regime, because their error increases significantly when their degree of exactness is not high enough. Consequently, methods with more gradually increasing rank profiles can compete with Gauss methods or can even outperform them for low number of sampling points.
This is demonstrated in the inset of Fig.~\ref{fig:cofcl}(d), which compares different methods for quadrature errors larger than \SI{0.1}{\%} and less then 100 sampling points.
In the present example, the equidistant step method (T$\times$T) allows to evaluate the PECD with relative error of \SI{1}{\%} with $72$ points. The Gauss product method (GL$\times$T) yields a relative error of about $0.1\,\%$ with only $50$ but has a higher error with more points, indicating that this could be a serendipitous effect of error cancellation. Even smaller quadrature errors using the same number of sampling points can potentially be achieved by optimizing the sampling densities of the azimuthal and polar integrals individually.

\subsection{High-rank, three angles: optimal control of ion yield circular dichroism}
\label{examples:cd}
\begin{figure*}
\includegraphics{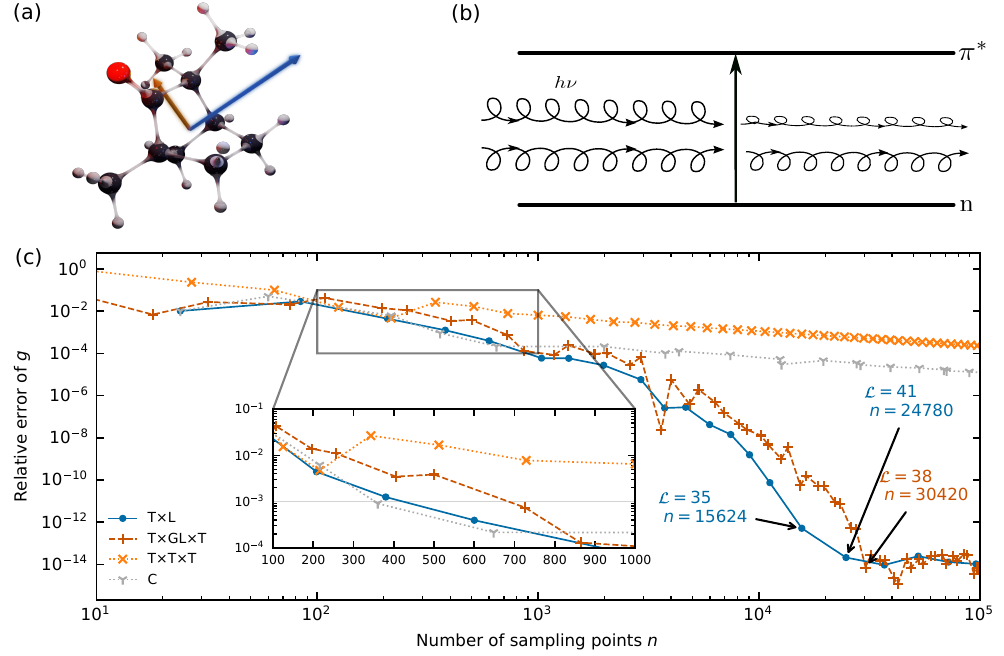}
\caption{
\textbf{(a)} Visualization of the fenchone molecule. The orange arrow indicates the electric dipole transition moment (scaled $\times 400$) and the blue arrow indicates the magnetic dipole transition moment (scaled $\times 4$). 
\textbf{(b)} Illustration of circular dichroism in fenchone. The system is treated via an effective two-level model corresponding to the electronic A band transition $n\rightarrow\pi^*$. The interaction of the chiral molecule with circular polarized (represented by helical arrows) light results in a difference in absorption.
\textbf{(c)} Relative quadrature error of the anisotropy factor $g$ calculated with Eq.~(\ref{eq:relative_quadrature_error}).
The inset displays a zoom for the region of relative quadrature errors around $0.1\%$. Methods are the composite trapezoid rule (T), Lebedev-Laikov quadrature (L), Gauss-Legendre quadrature (GL), and near-uniform \SO{3} coverings (C) from Ref.~\onlinecite{Karney2007}, with a cross indicating product methods.
The equidistant step method (T$\times$T$\times$T) uses the same number of points for all Euler angles.
}
\label{fig:CD_fenchone}
\end{figure*}
An example for a three-angle orientation average is absorption circular dichroism (CD) with polarization-shaped pulses, as discussed in Ref.~\onlinecite{Mondelo-Martell2022}. CD describes the difference in absorption of left- and right-circularly polarized light by a chiral molecule \cite{Nakanishi1994}, as schematically illustrated in Fig.~\ref{fig:CD_fenchone}(a) and (b). 
CD is quantified in terms of the anisotropy factor $g$ \cite{Mondelo-Martell2022},
\begin{align}
    g=\frac{I_\text{left}-I_\text{right}}{\frac{1}{2}\left(I_\text{left}+I_\text{right}\right)},
\end{align}
where $I_\text{left}$ and $I_\text{right}$ represent the population in the excited state for the two enantiomers after absorption. Anisotropy was enhanced via optimal control, leading to optimized driving fields with elliptical polarization and a DC component \cite{Mondelo-Martell2022_cor}.
Since neither the molecule nor the pulse has symmetries, the CD depends on all three Euler angles. In contrast to the previous case studies, the light-molecule interaction is non-perturbative due to the high intensity of the light. Thus, no ab initio prediction about the rank profile is possible and despite the molecules being uniformly distributed a high rank is to be expected.

A comparison of the quadrature error for different methods is provided in Fig.~\ref{fig:CD_fenchone}(c).
Once again, quadratures with non-zero degree displays super-exponential error scaling, outperforming the three-angle equidistant step method and the near-uniform spherical coverings.
The product quadrature combining the composite trapezoid and the Lebedev-Laikov quadrature (T$\times$L) shows the best convergence behaviour and achieves machine precision with the fewest points and degree $\acc=41$.
Due to the absence of symmetries in the physical setup, the quadrature error exhibits identical behavior regardless of whether the trapezoid method is applied to the first or third Euler angle.
The \SO{3} Gauss method from Ref.~\onlinecite{Graf2013}, though similar in behavior, is limited by its maximum degree of 14, which is insufficient for machine precision and thus omitted from Fig.~\ref{fig:CD_fenchone}(c).
The three-angle Gauss product method (T$\times$GL$\times$T) also reaches machine precision (with $\acc=38$) but requires $5640$ more points than the T$\times$L method due to its lower efficiency.
In principle, the T$\times$L method would converge to machine precision for a smaller $\acc$ (closer to $\acc=38$), also indicated by the shallow slope between the relative error at $\acc=35$ and $\acc=41$. However, no Lebedev-Laikov quadratures exist for degrees between $\acc=35$ and $\acc=41$.

Reaching machine precision requires about $25000$ points which is often impractical. Therefore, the inset in Fig.~\ref{fig:CD_fenchone}(c) shows a zoom on the area where the relative error is around \SI{0.1}{\%}.
In this range, the spherical coverings drop below a relative error of \SI{0.1}{\%} for the lowest number of sampling points, outperforming the T$\times$L method and vastly surpass the equidistant step method, which requires about two orders of magnitudes more points for the same error. Product methods built from the trapezoid rule and \unitsphere{2} coverings (data not shown) only slightly outperform the step method. In accordance with the results from Sec.~\ref{examples:cofcl}, this indicates that spherical coverings are particularly useful for integrands with high anisotropy.

\section{Python package}
\label{averaging:python_package}
Several of the quadrature methods discussed in this work are well-documented in the scientific literature or openly available from various online sources. For instance, Lebedev-Laikov quadrature points can be obtained from the original publications \cite{Lebedev1975,Lebedev1976,Lebedev1977,Lebedev1992,Lebedev1995,Lebedev1999} or from John Burkardt's homepage~\cite{BurkardtWebsite}, which also provides supplementary Fortran and Matlab code. Manuel Gräf's website \cite{GrafWebsite} offers sets of spherical Gauss quadratures and spherical Chebyshev quadratures on \unitsphere{2} and \SO{3}. A large collection of spherical designs on \unitsphere{2} and \unitsphere{3}, with the putative minimum number of points, is available from the website of Rob Womersley \cite{WomersleyWebsite}. Additionally, Charles Karney provides near-uniform coverings of orientation space on GitHub \cite{KarneyGithub}, and a broad collection of spherical designs and various types of spherical coverings can be downloaded from the homepage of Neil Sloane \cite{SloaneWebsite}. Efficient implementations of one-dimensional composite Newton-Cotes formulas or Gauss-Legendre sampling points and weights are also available via the NumPy and SciPy software libraries \cite{NumPy2020,SciPy2020}.

However, to the best of our knowledge, there is no common database for spherical quadrature methods. Although there have been some attempts within the open-source community to create such a resource, these efforts are often incomplete or no longer maintained. Specifically, we are not aware of any resources that facilitate the easy implementation of product quadratures. This lack of a centralized, reliable source complicates the process of selecting the optimal quadrature method for specific applications.

To address this gap, we have compiled a database of quadratures from most of the aforementioned sources, wrapped in a Python interface that allows users to easily access and implement these methods in their simulations. Published under the Mozilla Public License \cite{MPLv2} on GitHub \cite{PythonPackage}, the package is platform-independent and easily accessible. It can directly evaluate integrands in the form of Python functions or data arrays, and it also supports storing quadrature points and weights in files, allowing for integration with software written in other languages. For details on the implementation and examples, we refer to the documentation of the package \cite{PythonPackage}.

In addition to providing direct access to a variety of existing quadratures for different integration domains, the package allows individual quadrature methods to be used as building blocks to construct higher-dimensional product methods. This flexibility makes it a valuable resource for numerical integration on arbitrary one-dimensional intervals and multi-dimensional domains.
We encourage contributions from the community to maintain and expand this database.

\section{Conclusions}
We have conducted a comprehensive review of spherical quadrature methods suitable for the numerical evaluation of Euler angle integrals as encountered in the calculation of observables in gas or liquid phase experiments. We have identified five categories, derived from their mathematical and numerical properties: spherical Gauss quadrature, spherical Chebyshev methods, spherical coverings, product methods, and Monte-Carlo integration.
Two measures of performance have turned out to be useful: (i) the McLaren efficiency, useful for estimating the asymptotic numerical cost, and (ii) rank profiles, which describe how the quadrature error accumulates due to the individual anisotropic contributions of the integrand.
We have derived explicit guidelines for choosing the best quadrature method based on the rank profile of the integrand, illustrated by three state-of-the-art research examples from molecular physics. 
Specifically, we have provided a blueprint for estimating the rank profiles of angle-resolved photoelectron spectra based on the strength and symmetry of the light-molecule interaction.

When very high precision is desired, our convergence analysis of the examples has demonstrated that methods capable of exactly integrating individual terms of a series approximation of the integrand, i.e., spherical Gauss and Chebyshev quadratures as well as Gauss product methods, outperform near-uniform spherical coverings and equidistant step methods.
In particular, spherical Gauss methods such as the Lebedev-Laikov quadrature on \unitsphere{2} have conjectured optimal efficiency and consistently yield the lowest quadrature errors over a wide range of sampling points.
Thus, they are the best choice not only for low-rank integrands where exact integration is feasible, but also generally for applications in which the integrand can be expected to have smoothly decaying rank profiles. Given the lack of \SO{3} Gauss methods with degree $\acc>14$, the combination of the Lebedev-Laikov method with the composite trapezoid rule serves as a highly efficient alternative to Gauss quadratures for three-angle averages.

When a comparatively low precision is sufficient, 
Gauss quadratures are not necessarily the optimal choice, in particular, if the rank of the integrand is expected to be high. In such cases, near-uniform spherical coverings and even equidistant step methods can offer competitive performance with potentially lower computational costs.
Additionally, the examples we have studied indicate that product methods can potentially yield the smallest quadrature errors with a minimal number of sampling points if carefully tailored to the rank profiles of the integrand.

The examples suggest our guidelines to be valid for the different categories of quadratures regardless of the specific domain they are defined on. We thus expect our results to be applicable to numerical integration on higher-dimensional spherical domains and different geometries as well.

Future work should provide a more in-depth inspection of the rank profiles of different physical observables. Specifically, the notion of how symmetries in the light-molecule interaction influence their dependence on the first and third Euler angle could help to fully leverage the flexibility of product methods.

\section*{Acknowledgements}
We would like to thank Bar Ezra, Leon Kerber and Manel Mondelo-Martell for many helpful discussion. Financial support from Deutsche Forschungsgemeinschaft (DFG, German Research Foundation) --- project number 328961117 --- SFB 1319 ELCH is gratefully acknowledged.

\section*{Data availability statement}
The data that supports the findings of this study are available from the corresponding author upon reasonable request.
Quadrature methods used in this work are available on GitHub, reference number~\onlinecite{PythonPackage}.

\section*{Conflict of interest}
The authors declare no competing interests.

\section*{Author contributions}
A.B. carried out the literature review, the convergence analysis of photoionization of COFCl, and any analytical calculations. R.M.M.E. carried out the convergence analysis of optimal control of CD for fenchone. M.H. carried out the convergence analysis of multi-photon ionization of camphor. D.M.R. and C.P.K. planned and supervised the project. All authors contributed to the discussion of the results and the writing of the manuscript.

\appendix

\section{Efficiency of selected quadrature methods}
\label{averaging:appendix:efficiency_chart}
\begin{figure*}
\centering
\includegraphics[scale=1]{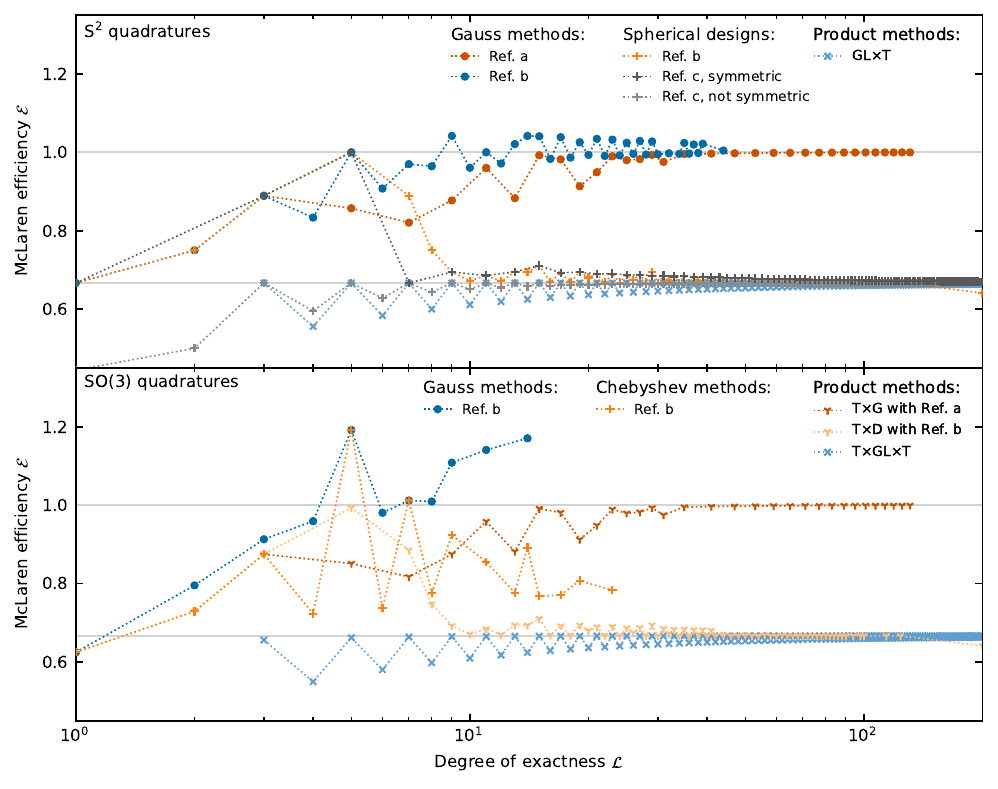}
\caption{
McLaren efficiency, Eq.~(\ref{eq:averaging:efficiency}), of the quadrature methods currently implemented in our software package \cite{PythonPackage}.
Methods are spherical Gauss quadrature (G), \unitsphere{2} spherical designs (D), the composite trapezoidal rule (T) and Gauss-Legendre quadrature (GL).
Reference~a corresponds to Refs.~\onlinecite{Lebedev1975,Lebedev1976,Lebedev1977,Lebedev1992,Lebedev1995,Lebedev1999}, Ref.~b corresponds to Ref.~\onlinecite{GrafWebsite}, Ref.~c corresponds to Ref.~\onlinecite{WomersleyWebsite}, which provides \unitsphere{2} designs with and without antipodal symmetry.
The gray lines indicate the conjectured optimal efficiency of spherical Gauss methods, $\eff=1$, and spherical Chebyshev methods, $\eff=2/3$.
}
\label{fig:efficiency}
\end{figure*}

We provide a chart of the McLaren efficiency of the spherical quadrature methods currently available in our software package \cite{PythonPackage} in Fig.~\ref{fig:efficiency}. This chart can be used to determine which quadrature method achieves a given degree with the fewest sampling points. Furthermore, this supplements our review of spherical quadratures in Sec.~\ref{averaging:overview}.
As the software package is expanded with additional quadrature methods, an updated version of this figure will be available in the documentation of the package \cite{PythonPackage}.

\section{Rank profiles on arbitrary domains}
\label{averaging:appendix:rank_profiles}
Rank profiles have been introduced by Ref.~\onlinecite{Eden1998} for quadratures on the two-dimensional unit sphere, \unitsphere{2}, and the rotation group, \SO{3}. This concept can be generalized to arbitrary domains. Here, we provide the general definition and concrete examples relevant for orientation averaging.

Consider the integral of a function $F(x)$ over the domain $\mathcal{D}$.
We represent the integrand in terms of a series expansion, the generalized form of Eq.~(\ref{eq:averaging:series_expansion}),
\begin{align*}
F(\orientation) = \sum_l \sum_{m_1} \dots \sum_{m_n} F_{l,m_1,\dots,m_n} \, p_{l,m_1,\dots,m_n}(\orientation)
\end{align*}
where $p_{l,m_1,m_2,\dots}(\orientation)$ are orthogonal basis functions on the domain $\mathcal{D}$ and $F_{l,m_1,m_2,\dots}$ are the expansion coefficients. We refer to $l$ as the rank of the basis function and explicitly allow for additional indices $m_1,\dots,m_n$, which may be used to further characterize basis functions of same rank. Examples would be the order $m$ of spherical harmonics, $Y_{l,m}(\theta,\phi)$, or the indices of Wigner D-matrices \cite{Abramowitz1964}.

Generalizing the definition from Ref.~\onlinecite{Eden1998}, we define the sampling moments on the domain $\mathcal{D}$,
\begin{align*}
\sigma_{l, m_1,\dots,m_n} = \frac{1}{V_{\mathcal{D}}} \int_{\mathcal{D}} p_{l,m_1,\dots,m_n}(\orientation) \mathrm{d}\orientation
\;,
\end{align*}
with $V_{\mathcal{D}}$ a normalization factor.
Correspondingly, we define the accumulated sampling moments,
\begin{align*}
\Sigma_l = \left( N_{m_1} \cdots N_{m_n} \sum_{m_1} \dots \sum_{m_n} |\sigma_{l, m_1,\dots,m_n}|^2 \right)^{1/2}
\;,
\end{align*}
and the accumulated expansion coefficients,
\begin{align*}
\Phi_l = \left( N_{m_1} \cdots N_{m_n} \sum_{m_1} \dots \sum_{m_n} |F_{l, m_1,\dots,m_n}|^2 \right)^{1/2}
\;,
\end{align*}
where $N_{m_i}$ denotes the number of terms in the summation over $m_i$.
For a given integration method, $\Sigma_l$ as a function of $l$ is the rank profile of the method and $\Phi_l$ as a function of $l$ is the rank profile of the integrand on the domain $\mathcal{D}$.
These can be used to analyze the numerical properties of quadratures methods on arbitrary domains with the procedure applied in Sec.~\ref{averaging:overview}.

Domains relevant for orientation averages in the form of the Euler angle integral, Eq.~(\ref{eq:averaging:euler_angle_integral}), are the unit circle, \unitsphere{1}, for the azimuthal integral, the interval $[-1, 1]$ for the polar integral, the two-dimensional unit sphere, \unitsphere{2}, for two-angle averages, and the rotation group \SO{3} for three-angle averages.
As orthogonal basis functions, we choose $e^{i l \phi}$ on \unitsphere{1}, Legendre Polynomials $P_l(z)$ on $[-1,1]$, spherical harmonics $Y_{l,m}(\theta,\phi)$ on \unitsphere{2} and Wigner D-matrix elements $D^{l}_{m_1,m_2}(\alpha,\beta,\gamma)$ on \SO{3}.
This yields the following definitions for the rank profiles:
\begin{widetext}
\begin{subnumcases}{\Sigma_l =}
\left|\sigma_l\right|
\quad&\text{with}\quad
$\displaystyle\sigma_l = \int_0^{2\pi} e^{i l \phi} \,\mathrm{d}\phi$
\quad \text{for quadratures on \unitsphere{1}}
\;,
\label{eq:averaging:appendix:method_rank_profile_phi}
\\
\left|\sigma_l\right|
\quad&\text{with}\quad
$\displaystyle\sigma_l = \int_{-1}^{1} P_l(z) \,\mathrm{d}z$
\quad \text{for quadratures \ensuremath{[-1,1]}}
\;,
\label{eq:averaging:appendix:method_rank_profile_theta}
\\
\left(\frac{1}{2l+1} \sum_{m} |\sigma_{l,m}|^2\right)^{1/2}
\quad&\text{with}\quad
$\displaystyle\sigma_{l,m} = \int Y_{l,m}(\Omega) \,\mathrm{d}\Omega$
\quad \text{for quadratures on \unitsphere{2} (see Ref.~\onlinecite{Eden1998})}
\;,
\label{eq:averaging:appendix:method_rank_profile_s2}
\\
\frac{1}{(2l+1)} \left(\sum_{m_1,m_2} |\sigma_{l,m_1,m_2}|^2\right)^{1/2}
\quad&\text{with}\quad
$\displaystyle\sigma_{l,m_1,m_2} = \int D^{l}_{m_1,m_2}(\orientation) \,\mathrm{d}\orientation$
\quad \text{for quadratures on \SO{3}}
\;,
\label{eq:averaging:appendix:method_rank_profile_so3}
\end{subnumcases}
\end{widetext}
and
\begin{widetext}
\begin{subnumcases}{\Phi_l =}
\left| F_l \right|
\quad &\text{on \unitsphere{1} and \ensuremath{[-1,1]}} \;,
\label{eq:averaging:appendix:integrand_rank_profile_1d}
\\
\left(\frac{1}{2l+1} \sum_{m} |F_{l,m}|^2\right)^{1/2}
\quad &\text{on \unitsphere{2} (see Ref.~\onlinecite{Eden1998})} \;,
\label{eq:averaging:appendix:integrand_rank_profile_s2}
\\
\frac{1}{(2l+1)} \left(\sum_{m_1,m_2} |F_{l, m_1, m_2}|^2\right)^{1/2}
\quad &\text{on \SO{3}} \;.
\label{eq:averaging:appendix:integrand_rank_profile_so3}
\end{subnumcases}
\end{widetext}
In particular, Eq.~(\ref{eq:averaging:appendix:method_rank_profile_s2}) has been used to calculate the rank profiles shown in Fig.~\ref{fig:method_rank_profiles}. To calculate the rank profile of the Euler angle distribution $P(\orientation)$ shown in Fig.~\ref{fig:cofcl}(c), we used Eq.~(\ref{eq:averaging:appendix:integrand_rank_profile_s2}).

\section{Rank profiles of selected quadrature methods}
\label{averaging:appendix:method_rank_profiles}
\begin{figure*}
\centering
\includegraphics[scale=1]{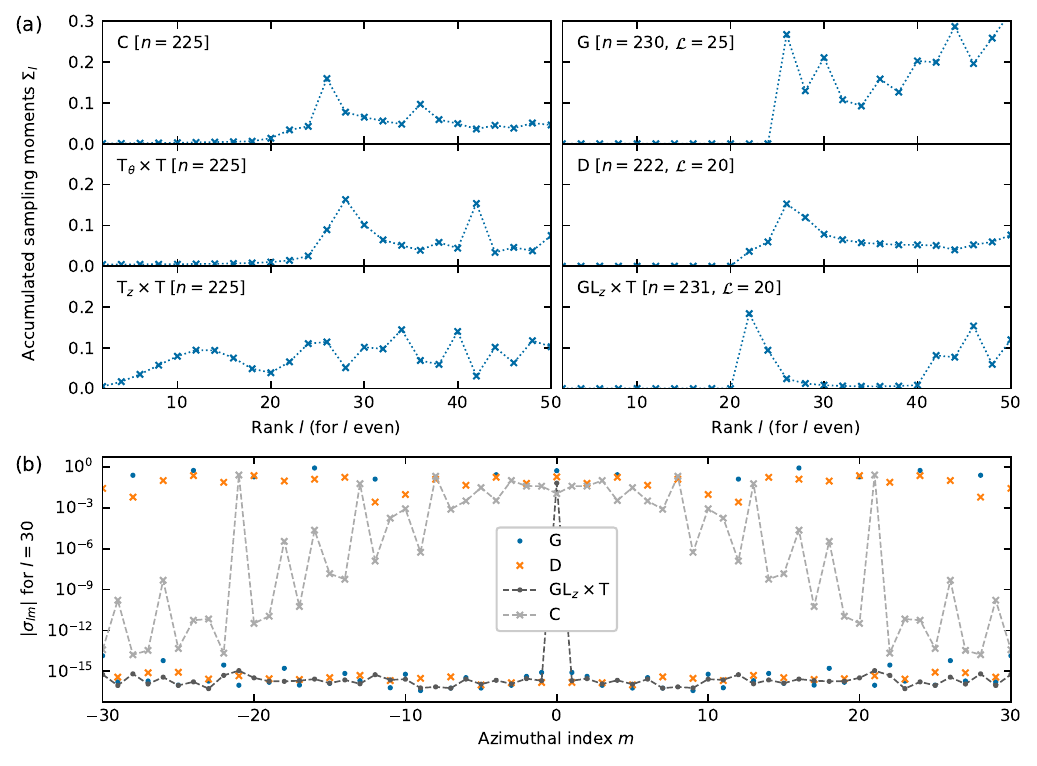}
\caption{
Rank profiles and sampling moments for selected two-angle quadrature methods.
\textbf{(a)} Rank profiles of \unitsphere{2} quadratures with zero degree (left column) and non-zero degree (right column).
The number of sampling points, $n$, and the degree, $\acc$, for which the corresponding rank profile is shown is denoted in braces behind the method name.
Methods are spherical Gauss quadrature from Ref.~\onlinecite{Lebedev1975,Lebedev1976,Lebedev1977,Lebedev1992,Lebedev1995,Lebedev1999} (G), spherical designs from Ref.~\onlinecite{WomersleyWebsite} (D), spherical covering with a Fibonacci grid (C), the composite trapezoid method (T) and Gauss-Legendre quadrature (GL).
\textbf{(b)} Sampling moments $\sigma_{lm}$ for $l=30$ for different $m$ for selected methods on \unitsphere{2}.
}
\label{fig:method_rank_profiles}
\end{figure*}

To supplement our review of spherical quadratures in Sec.~\ref{averaging:overview} we provide \unitsphere{2} rank profiles of selected methods representative for their respective categories in Fig.~\ref{fig:method_rank_profiles}.
To illustrate that the sampling moments of different quadrature methods can show significantly different dependence on the azimuthal indices $m_1,\dots,m_n$, we show \unitsphere{2} sampling moments $\sigma_{l,m}$ for selected methods in Fig.~\ref{fig:method_rank_profiles}(b). 

\section{Rank profiles of photoelectron angular distributions}
\label{averaging:appendix:rank_profile_of_pad}
In the perturbative expansion of photoelectron angular distributions (PADs), the orientation dependence arises in the form of products of rotation matrices \cite{Ritchie1976}. This can be used to derive an analytical expression for the rank profile of PADs in the perturbative regime.

To illustrate the procedure, we start from the expression in first-order perturbation theory for the anisotropy parameters of orientation-dependent laboratory-frame PADs considering ionization with arbitrarily polarized light \cite{Goetz2019b},
\begin{widetext}
\begin{align}
b^{(1)}_{LM}(\orientation)
&= \sum_{l_k,l_k'=0}^{l_{k,\mmax}} \sum_{m_k=-l_k}^{l_k} \sum_{m_k'=-l_k'}^{l_k'}
\sum_{\mu,\mu'=-1}^1 \sum_{\mu_0,\mu_0'=-1}^{1}
c(l_k, l_k', m_k, m_k', \mu, \mu', \mu_0, \mu_0', L, M) \times
\nonumber\\
&\hphantom{=}\times D^{1}_{\mu,-\mu_0}(\orientation) \, D^{1}_{-\mu',\mu_0'}(\orientation) \, D^{L}_{m_k'-m_k,-M}(\orientation)
\;.
\label{eq:averaging:appendix:orientation_dependent_1ph_betas}
\end{align}
\end{widetext}
The coefficients $c(l_k, l_k', m_k, m_k', \mu, \mu', \mu_0, \mu_0', L, M)$ collect the contributions from the transition dipole moments and from the time integrals and $l_{k,\mmax}$ denotes the highest partial wave populated in the partial wave expansion of the photoelectron's final state (see Ref.~\onlinecite{Goetz2019b}).
With the Euler angles $\orientation=(\alpha,\beta,\gamma)$ in the convention described in Sec.~\ref{averaging:euler_angle_integrals}, the Wigner D-matrix elements read
\begin{align*}
D^{l}_{m,m'}(\alpha, \beta, \gamma) =
e^{i m' \alpha} \, d^l_{m,m'}(\beta) \, e^{i m \gamma}
\;,
\end{align*}
where $d^l_{m,m'}(\beta)$ are the elements of the Wigner small d-matrix \cite{Edmonds1957,Zare1988}.
The last Wigner D-matrix element in Eq.~(\ref{eq:averaging:appendix:orientation_dependent_1ph_betas}) results from the transformation of the photoelectron momentum to the laboratory frame. The remaining rotation matrices originate from the frame transformation of the interaction Hamiltonian.
We repeatedly apply the contraction formula for products of Wigner D-matrices\cite{Edmonds1957},
\begin{widetext}
\begin{align}
D^{l_1}_{m_1, m_1'}(\orientation) D^{l_2}_{m_2, m_2'}(\orientation)
&=
\sum_{l=|l_1-l_2|}^{l_1+l_2} (2l+1) 
\begin{pmatrix}
l_1 && l_2 && l \\
m_1  &&  m_2  && -m_1-m_2
\end{pmatrix}
\begin{pmatrix}
l_1 && l_2 && l \\
m_1'  &&  m_2'  && -m_1'-m_2'
\end{pmatrix}
\times\nonumber\\
&\hphantom{=}\times
(-1)^{m_1 + m_2 + m_1' + m_2'}
D^{l}_{m_1+m_2, m_1'+m_2'}(\orientation)
\;,
\label{eq:averaging:appendix:wigner_D_contraction}
\end{align}
\end{widetext}
to bring the anisotropy parameters into the form of Eq.~(\ref{eq:averaging:series_expansion}).
The resulting expression reads,
\begin{align*}
b^{(1)}_{LM}(\orientation) = \sum_{l=L-2}^{L+2} \sum_{m=-l}^l F^{(1) LM}_{l m M} D^{l}_{m,M}(\orientation)
\end{align*}
with
\begin{widetext}
\begin{align}
F^{(1) LM}_{l m M}
&=
\sum_{l_k,l_k'=0}^{l_{k,\mmax}} \sum_{m_k=-l_k}^{l_k} \sum_{m_k'=-l_k'}^{l_k'}
\sum_{\mu,\mu'=-1}^1 \sum_{\mu_0,\mu_0'=-1}^{1}
c(l_k, l_k', m_k, m_k', \mu, \mu', \mu_0, \mu_0', L, M)
\sum_{j=0}^{2} (2j+1) (2l+1)
\times\nonumber\\
&\hphantom{=}\times
\begin{pmatrix}
1 && 1 && j \\
\mu  &&  -\mu'  && \mu'-\mu
\end{pmatrix}
\begin{pmatrix}
1 && 1 && j \\
-\mu_0  &&  \mu_0'  && \mu_0-\mu_0'
\end{pmatrix}
\begin{pmatrix}
j && L && l \\
\mu'-\mu  &&  m_k'-m_k  && -m
\end{pmatrix}
\begin{pmatrix}
j && L && l \\
0  &&  -M  && M
\end{pmatrix}
\times\nonumber\\
&\hphantom{=}\times
(-1)^{m_k + m_k'+ M}
\delta_{m, \mu-\mu' + m_k-m_k'}
\;.
\label{eq:averaging:appendix:rank_profile_of_1ph_pad}
\end{align}
\end{widetext}
The $F^{(1) LM}_{l m M}$ constitute the rank profile of the individual anisotropy parameters of a PAD resulting from a one-photon process.
The Wigner 3j-symbols restrict the summation over $j$ to $j\leq 2$, which in turn restricts the maximum rank of the expansion coefficients to $l\leq L+2$.

Equation~(\ref{eq:averaging:appendix:rank_profile_of_1ph_pad}) can be straightforwardly extended to higher orders.
Each application of the interaction operator, i.e. each photon involved in the ionization, contributes a product of two additional rotation matrices to the final expression of the PAD \cite{Goetz2017,Goetz2019,Goetz2019b}.
By repeated contraction of the Wigner D-matrices with Eq.~(\ref{eq:averaging:appendix:wigner_D_contraction}), we can rewrite the $n_{\mathrm{ph}}$-photon PAD from Eq.~(\ref{eq:averaging:pad_expansion}) as
\begin{widetext}
\begin{align}
\PAD(\Omega_k, \orientation) =
\sum_{L=0}^{2 l_{k,\mmax}} \sum_{M=-L}^{L} \sum_{l=0}^{L+2n_{\mathrm{ph}}} \sum_{m=-l}^{l} F^{(n_{\mathrm{ph}}) LM}_{l m M} \, D^{l}_{m,M}(\orientation) \, Y_{LM}(\Omega_k)
\;,
\label{eq:averaging:appendix:rank_profile_of_nph_pad}
\end{align}
\end{widetext}
As for the one-photon expression, the $F^{(n_{\mathrm{ph}}) LM}_{l m M}$ contain a product of Wigner 3j-symbols that for each $L$ restrict their rank to $l \leq L+2n_{\mathrm{ph}}$.
The maximum rank of each anisotropy parameter is thus determined by the number of molecule-photon interactions.

The number of anisotropy parameters, i.e. the summation over $L$, is bounded by the angular momentum of the photoelectron, encoded in the partial wave expansion of the photoelectron's final state according to $L \leq 2 l_{k, \mmax}$ \cite{Ritchie1976,Goetz2017}.
The latter is bounded by the angular momentum of the molecule's initial state and the number of molecule-photon interactions, giving $l_{k, \mmax} \leq l_{0, \mmax} + n_{\mathrm{ph}}$ for electric dipole transitions with circularly polarized light, where $l_{0, \mmax}$ is the largest angular momentum quantum number of the initial state.
In total, we can formulate the upper bound for the maximum rank of the $n_{\mathrm{ph}}$-photon PAD as
\begin{align*}
l_{\mmax}
&\leq L_{\mmax} + 2 n_{\mathrm{ph}}
\\
&\leq 2 l_{k, \mmax} + 2 n_{\mathrm{ph}}
\\
&\leq 2 l_{0, \mmax} + 4 n_{\mathrm{ph}}
\;.
\end{align*}
corresponding to Eq.~(\ref{eq:averaging:max_rank_of_pad}) in the main text.
Note that the values of $l_{k, \mmax}$ or $l_{0, \mmax}$ depend on the convergence of the expansion of the molecular states in terms of the corresponding angular momentum basis states. This expansion may not have an exact cutoff. Instead $l_{k, \mmax}$ and $l_{0, \mmax}$ may be defined by the expansion coefficients becoming smaller than a reference value (e.g. a given numerical precision). In this case, the maximum rank of the PAD depends on this reference value.

The exact relation between the angular momentum of the initial state and the angular momentum of the final state is determined by the symmetry of the molecular states, which is ultimately connected to the symmetry group of the molecule. The molecular symmetry may thus be used to find even tighter bounds for the maximum rank of the PAD.
Furthermore, if the initial state is a thermal state of finite temperature, the initial angular momentum, and hence the maximum rank of the resulting PADs, is higher compared to the situation that the molecule initially is in an low-energy eigenstate.

\section{Reduced orientation averages of photoelectron angular distributions}
\label{averaging:appendix:simplified_averages}
The behavior of the $F^{(n_{\mathrm{ph}}) LM}_{J_n m_n M}$ from Eq.~(\ref{eq:averaging:appendix:rank_profile_of_nph_pad}) for different $M$ and $m_n$ describes the dependence of the PAD on the first and third Euler angles, $\alpha$ and $\gamma$, respectively. In the following, we illustrate, how this can be used to simplify the orientation averaging procedure.

\subsection{Cylindrically symmetric molecules}
\label{averaging:appendix:averaging_cylindrical_molecules}
The allowed values of $m_n$ arise from the projections of the partial wave expansion of the final state of the photoelectron onto the $z$-axes of the laboratory-frame and of the molecular-frame. Higher azimuthal anisotropy of the partial wave expansion yields stronger dependence of the PAD on $\gamma$. Which partial waves are populated during the ionization depends on the selection rules between the molecular states. Since these selection rules arise from the molecular symmetry, the dependence of PADs on the third Euler angle can thus already be deduced from the symmetry group of the molecule. In the limit that the molecule is cylindrically symmetric, e.g.~for linear molecules, the PAD does not depend on $\gamma$. This simplifies the orientation average of symmetric molecules substantially.

\subsection{Cylindrically symmetric pulses}
\label{averaging:appendix:averaging_cylindrical_pads}
The dependence of the PAD on the first Euler angle, $\alpha$, is directly connected to the dependence of the PAD on the azimuthal angle of the photoelectron momentum, $\phi_k$, encoded in $M$. Hence, PADs that are cylindrically symmetric around the laboratory $z$-axis, i.e.~only contain terms with $M=0$, do not depend on $\alpha$.
This is the case for ionization with circularly polarized pulses that contain enough cycles to be well approximated as cylindrically symmetric in the electric dipole approximation. With the help of Eq.~(\ref{eq:averaging:appendix:rank_profile_of_1ph_pad}), it can be shown that this condition is fulfilled if the PAD can be described in the rotating wave approximation (RWA). However, note that interferences of different ionization pathways can break the cylindrical symmetry of a PAD \cite{Goetz2019}.

For cylindrically symmetric PADs, the first Euler rotation around the pulse propagation direction results in a complex phase factor, $\exp(-iM\alpha)$, which corresponds to a rotation of the PAD around the propagation direction of the pulse. Averaging over $\alpha$ thus only eliminates all anisotropy parameters with $M\neq0$ without additional effects on the PAD. Hence, it is sufficient to calculate the PAD for $\alpha=0$ and to perform a two-angle orientation average, followed by an integration of the averaged PAD over $\phi_k$ to recover the cylindrical symmetry.
This strategy can often be exploited to reduce the numerical effort in calculating photoelectron spectra in the electric dipole approximation, see e.g.~Refs.~\onlinecite{Artemyev2015,Tia2017,Muller2018,Demekhin2018,Demekhin2019,Muller2020,Tikhonov2022} and the examples given in Secs.~\ref{examples:rempi} and \ref{examples:cofcl}.
However, care needs to be taken when working with very short (i.e. few-cycle) pulses, which do permit to use the RWA and hence cannot be approximated as cylindrically symmetric. For such pulses, the $\alpha$-dependence of the PAD becomes increasing complicated and the averaged PAD starts to loose its cylindrical symmetry with decreasing pulse duration.
The same is true for elliptically polarized pulses with increasing eccentricity (as in the example from Sec.~\ref{examples:cd}) or in the presence of interfering ionization pathways, which are known to break the cylindrical symmetry of the PAD by giving rise to anisotropy parameters with $M\neq0$ \cite{Goetz2019,Goetz2019b} even for circularly polarized pulses.
Nevertheless, one can reduce the numerical cost of the orientation average in these situations by choosing a product quadrature method that uses less sampling points for $\alpha$ than for the other Euler angles.

\section{Euler angle distributions of rotational wavepackets}
\label{averaging:appendix:euler_angle_distribution_and_rotational_states}
The rotational state describes the orientational distribution of a molecule and can thus be used to construct the probability density for an anisotropic orientation average. Hence, the maximum rank of Euler angle distributions is inherently connected to the maximum orbital angular momentum of the molecule, as we illustrate in the following.

Let $\ket{\psi_{\mrot}}$ be the rotational state of a quantum mechanical rigid rotor, which we represent in terms of the eigenstates of a symmetric top molecule,
\begin{align}
\ket{\psi_{\mrot}} = \sum_{JKM} c_{JKM} \ket{JKM}
\;,
\label{eq:averaging:appendix:rotational_state_expansion}
\end{align}
with complex expansion coefficients $c_{JKM}$.
The parameterization of these eigenstates in terms of Euler angles is given by Wigner D-matrix elements, $\braket{\orientation | JKM} = D^{J}_{KM}(\orientation)$. \cite{Zare1988,Bunker2012}
We obtain the Euler angle distribution for the orientation average with Eq.~(\ref{eq:averaging:euler_angle_integral}) as $P(\orientation) = ||\ket{\psi_{\mrot}}||^2$. Inserting the expansion from Eq.~(\ref{eq:averaging:appendix:rotational_state_expansion}) and using the contraction property of Wigner D-matrix elements, Eq.~(\ref{eq:averaging:appendix:wigner_D_contraction}), we can write
\begin{widetext}
\begin{align*}
P(\orientation)
&= \sum_{JKM} \sum_{J'K'M'} c_{JKM} c_{J'K'M'}^* \, D^{J}_{KM}(\orientation) D^{J'}_{K'M'}(\orientation)^*
\\
&= \sum_{l}
\sum_{JKM} \sum_{J'K'M'} c_{JKM} c_{J'K'M'}^* \,
(-1)^{K-M} (2l+1)
\begin{pmatrix}
J && J' && l \\
K  &&  -K'  && K'-K
\end{pmatrix}
\begin{pmatrix}
J && J' && l \\
M  &&  -M'  && M'-M
\end{pmatrix}
\times
\nonumber\\
&\hphantom{=} \times D^{l}_{K-K',M-M'}(\orientation)
\end{align*}
\end{widetext}
The Wigner 3j-symbols restrict the summation over $l$ to $l\leq 2J_{\mmax}$, where $J_{\mmax}$ is the maximum rotational quantum number contributing to the rotational state $\ket{\psi_{\mrot}}$ in Eq.~(\ref{eq:averaging:appendix:rotational_state_expansion}).
The prefactor in front of the Wigner D-matrix elements in the final expression constitutes the rank profile of $P(\orientation)$ as defined in Eq.~(\ref{eq:averaging:appendix:integrand_rank_profile_so3}) or equivalently in Eq.~(\ref{eq:averaging:rank_profile_integrand_three_angle}).
Hence, the maximum rank of an Euler angle distribution constructed from a rotational quantum state is given by twice the maximum rotational quantum number of the molecule and hence has a direct relation to the orbital angular momentum of the molecule.

\section*{References}
\bibliography{averaging_methods,averaging_examples}

\end{document}